# The precession of vortex-beams in a rotating uniaxial crystal

T. Fadeyeva, A. Rubass, B. Sokolenko, A. Volyar

*Physical Department, Taurida National University, Simferopol 95007, Crimea Ukraine*
*Corresponding author: volyar@crimea.edu

**Abstract** We consider theoretically and experimentally the precession of the optical vortex in a singular beam propagating nearly perpendicular to the crystal optical axis, the beam and the crystal axis rotating with different angular velocities. The precession is a result of different scales along the major axes in a uniaxial crystal. Also we analyze the properties of the optical reducer consisting of two rotating crystals and a wave retarder between them permitting us to control the form of the vortex trajectory.

Unique properties of birefringent crystals to transform Gaussian beam into a singular one [1] and control parameters of optical vortices when propagating the beam along or nearly along [2] the crystal optical axis find a use in different devices for trapping, transportation and angular orientation of microparticles [3]. While transmitting the Gaussian beam perpendicular to the crystal optical axis, the cross-section of the extraordinary beam is elliptically deformed [4,5]. Naturally, the elliptical deformation transforms a fine vortex structure in a singular beam and ultimately may come to unexpected phenomena.

In the Letter we focus our attention on movement of optical vortices in a singular beam transmitting through a rotating uniaxial crystal, the beam propagating nearly perpendicular to the crystal optical axis.

**1.** We consider theoretically a tilted circularly polarized monochromatic singular beam bearing the first order optical vortex (so-called a vortex-beam) transmitting at a small angle $\alpha_o \ll 1$ ($\sin\alpha_o \approx \alpha_o$, $\cos\alpha_o \approx 1 - \alpha_o^2/2$) to the $z$-axis of the crystal referent frame $(x, y, z)$ shown in Fig.1. The optical vortex is displaced on the beam cross-section $z=0$ at the distance $\rho_0$. In the theoretical aspect, the anisotropic medium of the crystal is assumed to be unbounded one with a permittivity tensor in a diagonal form: $\hat{\varepsilon} = diag\, \varepsilon_{ij}$ so that $\varepsilon_{11} = \varepsilon_{33} = \varepsilon_1$, $\varepsilon_{22} = \varepsilon_2$ in the coordinates $(x, y, z)$, the crystal optical axis **C** directing perpendicular to the $z$-axis. We restrict ourselves to a case of the crystal rotating around the $z$-axis with the angular velocity $\Omega_C \ll \omega$ (i.e. the rotation angle $\psi = \Omega_C t$, t is a time time, $\omega$ is a light wave frequency), the beam with a shifted vortex can rotate also around the beam $z'$-axis with the angular velocity $\Omega_V \ll \omega$ ($\phi = \Omega_V t$) while the axis of the ordinary beam and the observer are motionless. The observer is lodged in the $x_1, y_1, z_1$ referent frame whereas the beam is defined in the $x', y', z'$ coordinates. In general case, the beam has an elliptic cross-section [5]. Its angular position is defined by Euler' angles: $\alpha_o$ - a nutation angle $\psi$ - a precession angle and $\phi$ - an angle of intrinsic rotation shown in Fig.1a. The $y'$ axis is directed along one of the ellipse axes of the beam cross-section.

A particular solution for the vortex-beam to the paraxial wave equation for a complex amplitude $\widetilde{\mathbf{E}}_\perp = \{\widetilde{E}_x, \widetilde{E}_y\}$ of the electric field

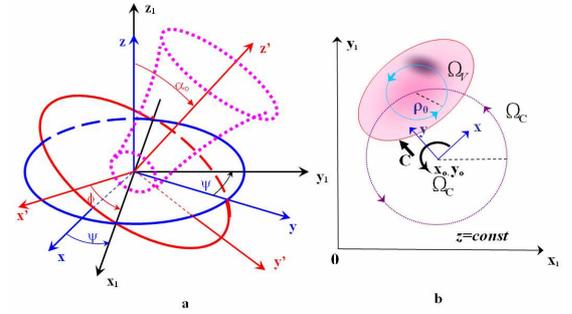

Fig.1 (a) The elliptic beam in the crystal and its Euler' angles; (b) a sketchy representation of the beam and the crystal rotation (**C** is the crystal optical axis).

$\mathbf{E}_\perp = \widetilde{\mathbf{E}}_\perp(x,y,z)\exp(-ik_j z)$ ($k_j$ stands for a wavenumber, $j = 1, 2$) [6] can be written as:

$$E_x = \left(x - \alpha_o z\cos\psi + i\xi(y - \alpha_o z\sin\psi) - a w_o \sigma_o e^{-i\xi(\psi-\varphi)}\right)*$$
$$*\exp\left[-\left(X_o^2 + Y_o^2\right)/w_o^2\sigma_o - f_o - ik_1 z\right]/w_0\sigma^2 \quad (1)$$

$$E_y = \frac{i}{\sqrt{\sigma_x \sigma_y}}\left(\frac{x - \alpha_x z\cos\psi}{w_e \sigma_x} + i\xi\frac{y - \alpha_y z\sin\psi}{w_e \sigma_y} - a e^{-i\xi(\psi-\varphi)}\right)*$$
$$*\exp\left[-\frac{X_e^2}{w_e^2\sigma_x} - \frac{Y_e^2}{w_e^2\sigma_y} - f_e - ik_2 z\right] \quad (2)$$

where, $\sigma_o = 1 - iz/z_o$, $\sigma_x = 1 - iz/z_x$, $\sigma_y = 1 - iz/z_y$, $z_o = k_1 w_o^2/2$, $z_x = k_2 w_e^2/2$, $z_y = k_2 w_e^2 n_1^2/2n_2^2$, $X_o = x + i\alpha_o z_o \cos\psi$, $Y_o = y + i\alpha_o z_o \sin\psi$, $X_e = x + i\alpha_x z_x \cos\psi$, $Y_e = y + i\alpha_y z_y \sin\psi$, $f_e = \alpha_x^2 k_2 z_x \cos^2\psi/2 + \alpha_y^2 k_2 z_y \sin^2\psi/2$, $f_o = \alpha_o^2 k_1 z_o/2$, $k_1 = n_1 k_0$, $k_2 = n_2 k_0$, $n_1 = \sqrt{\varepsilon_1}$, $n_2 = \sqrt{\varepsilon_2}$, $k_0$ stands for a wavenumber in free space, $a$ is a coefficient of the vortex displacement, $\xi = \pm 1$, $w_o$ and $w_e$ are the radii of the ordinary $E_x$ and extraordinary $E_y$ beam waists at the plane $z = 0$. Besides, our requirement is the amplitudes and waist radii $w_o$ and $w_e$ in the $E_x$ and $E_y$ components to be equal to each other at the plane $z=0$: $w_o = w_e = w_0$. From whence we come to $\alpha_x z_x = \alpha_y z_y = \alpha_o z_o$, $f_o = f_e$ and, consequently, $n_2\alpha_x = n_1\alpha_o$, $n_1\alpha_y = n_2\alpha_o$. The different values of the parameters $z_x$ and $z_y$ associated with different scales along the major axes in the crystal entails the elliptical deformation of the extraordinary $E_y$ beam cross-section. Since the beam has a circular cross-section at the $z=0$ plane, the direction of the beam deformation at the arbitrary plane $z=const$ will always follow the crystal optical axis **C** (variation of the angle $\psi$) when rotating the crystal despite the beam by itself can rotate around its axis (variation of the angle $\phi$).

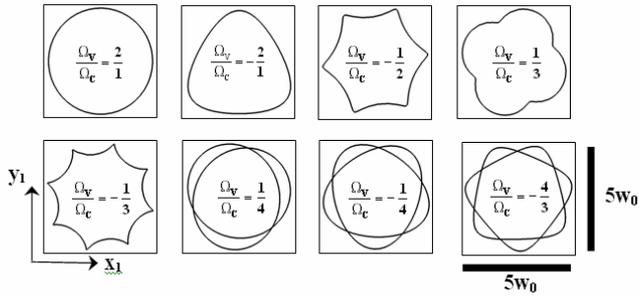

Fig.2 Major types of the vortex trajectories at the plane z=2cm in the crystal with $n_1 = 2.4$, $n_2 = 2.2$ and $w_o = 10\,\mu m$, $a = 0.7$

**2.** Let consider the precession of a shifted optical vortex recoded by the motionless observer in the $x_1, y_1, z_1$ referent frame when rotating the crystal optical axis and the shifted optical vortex (i.e. the angles $\psi$ and $\phi$ change). The situation is illustrated by Fig.1b. Notice that the vortex movement in eq. (2) is defined by two independent parameters $a$ and $\alpha_o$ that responsible for two independent rotations: a precession of the vortex around some axis inside the beam and a precession of the beam as a whole around some axis inside the crystal. The vortex position is defined as $\operatorname{Re} E_j(x,y,z) = \operatorname{Im} E_j(x,y,z) = 0$ ($j = x, y$). Since we take an interest in behavior of the optical vortex in the $E_y$ extraordinary beam we assume that the polarizer placed right after the crystal rotates together with the crystal letting pass through it only the $E_y$ component.

Thus, taking into account the relations between the beam coordinates in the referent frame of the crystal $x, y, z$ and the observer $x_1, y_1, z_1$:

$x = x_1 \cos\psi - y_1 \sin\psi$,
$y = x_1 \sin\psi + y_1 \cos\psi$ we find the form of the vortex trajectory in the extraordinary beam relative to a motionless coordinate system:

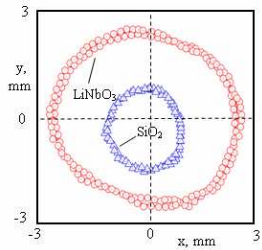

Fig.3 The precession of the shifted vortex

$$x_1 = x_0 + (T_1 - \alpha_o z\, B)\cos 2\psi + T_2 \sin 2\psi, \quad (4)$$

$$y_1 = y_0 - T_2 \cos 2\psi + (T_1 + \alpha_o z\, B)\sin 2\psi, \quad (5)$$

where $T_1 = A[(Z_x + Z_y)\cos\phi - \xi(1 - Z_x Z_y)\sin\phi]$,
$T_2 = A[(Z_x + Z_y)\sin\phi + \xi(1 - Z_x Z_y)\cos\phi]$, $Z_y = z/z_y$,
$Z_x = z/z_x$, $x_0 = \dfrac{aw_0}{2}\left[-\xi(Z_x + Z_y)\cos\phi + \dfrac{2 + Z_y^2 + Z_x^2}{1 + Z_x Z_y}\sin\phi\right]$,

$y_0 = \dfrac{aw_0}{2}\left[\dfrac{2 + Z_y^2 + Z_x^2}{1 + Z_x Z_y}\cos\phi + \xi(Z_x + Z_y)\sin\phi\right] - K$,

$K = \alpha_o z\,\dfrac{n_1^2 + n_2^2}{2n_1 n_2}$, $A = \dfrac{aw_0(Z_y - Z_x)}{2(1 + Z_x Z_y)}$, $B = \dfrac{n_2^2 - n_1^2}{2n_1 n_2}$.

The above equations represent a set of complex trajectories on the plane $z = const$ shown in Fig.2. Each trajectory is characterized by a ratio of the frequencies: $\Omega_V/\Omega_C$. Provided that this ratio represents a simple fraction $\Omega_V/\Omega_C = p/q$, the lines of the trajectories are closed. A circular trajectory corresponds to the case $p/q = n$ ($n = 0,1,2,...$). The sign (–) in Fig.2 indicates a counter-rotation of the crystal and the beam. In case the beam is motionless ($\Omega_V = 0$) the vortex precesses with the frequency $\Omega_V = 2\Omega_C$ along a circular trajectory with a radius $r_p = A\sqrt{(1 + Z_x^2)(1 + Z_y^2)}$. For example, for the *LiNbO3* crystal with $n_1 = 2.3$, $n_2 = 2.2$, the crystal length $z = 2\,cm$, $a = 1$ and the beam waist $w_o = 10\,\mu m$ the precession radius is $r_p \approx 4\,\mu m$ against a background of the beam spot with the radius $w \approx 20\,\mu m$.

For the experiment, we used the experimental set-up described in detail in Ref. [2]. The crystal was positioned in a rotary table permitting us to rotate the crystal in three orthogonal planes. The crystal optical axis was directed perpendicular to the vortex-beam axis. The vortex in the beam is displaced at the crystal input at the distance $x_0 = 0.7\,w_o$. The crystal is rotated around the $z$-axis to within $0.15^o$ while the beam is motionless. The slant of the parallel verges of the crystal does not exceed $0.07^o$. The deviation of the trajectory induced by this slant was taken into account by the computer processing of the experimental results. The crystal length was about 2 cm. The circularly polarized beam from the He-Ne laser carries over the optical vortex with the topological charges $l = \pm 1$. The $20^\times$ microobjective projected the beam spots at the output face of the crystal onto the input pupil of the CCD camera. The vortex trajectories shown in Fig.3 were plotted for two types of the crystals: *SiO2* and *LiNbO3*. A binary-ring trajectory of the vortex is caused by a faint slant of the crystal verges. Within the error of the experiment the trajectories have a circular form and do not depend on a sign of the vortex topological charge.

**3.** In case the beam with a centered vortex ($a = 0$, $\Omega_V = 0$, $\phi = 0$) is tilted at a small angle $\alpha_o \neq 0$, rotation of the crystal entails rotation of the extraordinary beam along a circular trajectory (see eqs (4, 5)), one revolution of the crystal axis causing two revolutions of the vortex-beam. However the beam is rotated not around a position of the ordinary beam $x = -\alpha_o z$, $y = 0$ (as it could be expected) but around the other point: $x_0 = -K$, $y_0 = 0$. The radius of the rotation is $r_0 = \alpha_o z B$.

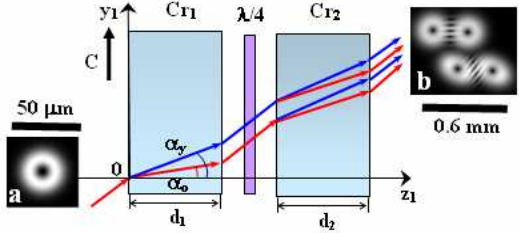

Fig.4

Fig.4 Sketchy representation of the optical reducer: (a), (b) vortex-beam at the input and output of the system (the computer simulation is performed for $n_1 = 2.4, n_2 = 2.2, d_1 = 2\,cm, d_2 = 2.8\,cm,$ $w_0 = 15\,\mu m, \alpha_o = 0.12\,rad, \Omega_1/\Omega_2 = 1/3$)

We will focus our attention on the more general case that is of interest for practical applications in particular for the devises of trapping and transportation of microparticles.

Let us consider two crystals positioned in series along the $z$-axis. A quarter-wave retarder is placed between the crystals and rotates synchronically with the first crystal (Fig.4). It supports a circular polarization at the input of the second crystal. A circular polarized beam with a centered vortex enters inside the first crystal at the plane $z=0$ at the angle $\alpha_o$. We choose the angle $\alpha_o$ in such a way that the ordinary and extraordinary beams are separately observed [2] at the output of the second crystal at the plane $z = d_1 + d_2$. The crystals can rotate independently at the angles $\psi_1 = \Omega_1 t$ and $\psi_2 = \Omega_2 t$. At the output of the system we can observe four beams: the *o-o, e-o, o-e* and *e-e* beams. The *o-o* beam passes through the crystal without rotation (it does not mark in Fig.5). The *e-o* beam originates from the extraordinary beam in the first crystal being the ordinary one in the second crystal. It traces the circular trajectory controlled by the angle $\psi_1$. The *o-e* beam arisen from the ordinary beam in the first crystal is controlled by the angle $\psi_2$ and moves also along the circular trajectory. The intricate trajectory of the *e-e* beam is controlled by two $\psi_1 = \Omega_1 t$ and $\psi_2 = \Omega_2 t$ angles. Its shape depends on a ratio of the frequencies $\Omega_1/\Omega_2$. Such a system enables us to control the form of the vortex trajectory and reduces angular velocities of the vortices. We called it the *optical reducer*. The computer simulation of the vortex-beam precession is shown in Fig.5.

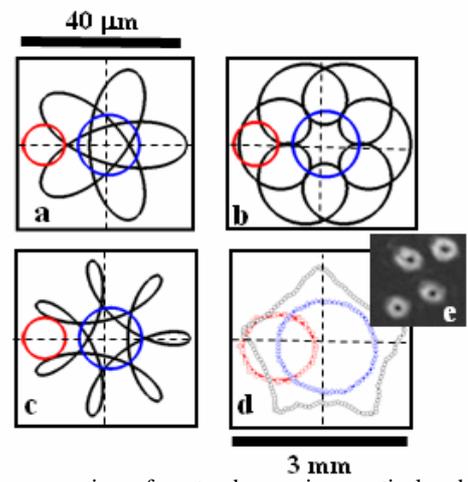

Fig.5 The precession of vortex-beams in a optical reducer with $n_1 = 2.3$, $n_2 = 2.2$, $d_1 = d_2 = 2\,cm$, $\alpha_o = 0.12\,rad$ for the ratio $\Omega_1/\Omega_2$: a) -2/3, b) -2/5, c) 1/7 (the computer simulation) d) the experiment $\Omega_1/\Omega_2 = 1/3$, e) the beams at the system output; red - *o-e*, blue – *e-o*, black – *e-e* beams.

For the experiment we used two *LiNbO₃* crystals with the same lengths $d_1 = d_2 = 2\,cm$ so that the slants of the parallel verges of the crystals were lesser than $0.1^o$. The quarter wave retarder placed between the crystals can transform the linear polarized light into the circular polarized one within a range of the inclination angles $\Delta\alpha = 6^o - 8^o$ with the efficiency about 93%. The distance between the crystals was about 1 cm. The experimental results shown in Fig.5d,e are illustrated by the typical circumferences (for the *o-e* and *e-o* beams) and the star (for the *e-e* beam) for the frequency ratio $\Omega_1/\Omega_2 = -1/3$.

Thus, we have theoretically and experimentally considered the precession process of the optical vortex stimulated by the crystal and the beam rotation. We have also analyzed the system of two rotating crystal (the *optical reducer*) that enable to control the form of the vortex trajectories.

Authors are grateful to E. Abramochkin and C. Alexeyev for useful discussion and to Yu. Egorov for the help in the experiments.